\documentclass[12pt,a4paper]{article}
\usepackage{amsmath}
\usepackage{amsfonts}
\usepackage{amssymb}
\usepackage{a4wide}
\newcommand{\initiate}{\setcounter{equation}{0}}
\def\a#1{\bar{#1}}                              
\def\ad{\mbox{ad}\,}                            

\def\c#1{{\cal #1}}                             
\def\Dirac{{\raise0.09em\hbox{/}}\kern-0.69em D}
\def\ep{i\epsilon}
\def\epr{\epsilon}
\def\kbar{{\mathchar'26\mkern-9muk}}            
\def\lesssim{\mathrel{\hbox{\rlap{\hbox{\lower4pt\hbox{$\sim$}}}\hbox{$<$}}}}
\def\ep{i\epsilon}
\def\p{\partial}                                
\def\tfrac #1#2{\textstyle{\frac{#1}{#2}}}      
\def\t#1{\tilde #1}                             
\def\h#1{\hat{#1}}                              
\def\tr{\mbox{Tr}\,}                            

\def\vev#1{\langle #1 \rangle}                  
\def\wm{\mathbin{*}}

\def\k{\kern-.1em\mathbin{,}\kern-.1em}
\def\hk{\kern.12em\raise-1em\hbox{$\hat{\raise1em\hbox{,}}$}\kern.12em}

\def\hI{{\hat I}}
\def\cA{{\bf A}}
\begin{document}
\title{The Energy-momentum of a Poisson structure}
\vskip25pt
\author{
        M. Buri\'c              $^{1}$\thanks{majab@phy.bg.ac.yu}
\ \
\\[10pt]
        J. Madore               $^{2}$\thanks{madore@th.u-psud.fr}
\ \
        G. Zoupanos             $^{3}$\thanks{zoupanos@cern.ch}
\\[25pt]
        $\strut^{1}$Faculty of Physics
\\               University of Belgrade, P.O. Box  368
\\               SR-11001 Belgrade
\\[10pt]
        $\strut^{2}$Laboratoire de Physique Th\'eorique
\\               Universit\'e de Paris-Sud, B\^atiment 211
\\               F-91405 Orsay
\\[10pt]
        $\strut^{3}$Physics Department
\\               National Technical University, Zografou Campus,
\\               GR-15780 Zografou, Athens
\\
}
\date{}
\maketitle

\parskip 10pt plus2pt minus2pt
\parindent 0pt

\vskip25pt

\abstract{Consider the quasi-commutative approximation to a
  noncommutative geometry. It is shown that there is a natural map
  from the resulting Poisson structure to the Riemann curvature of a
  metric. This map is applied to the study of high-frequency
  gravitational radiation. In classical gravity in the WKB
  approximation there are two results of interest, a dispersion
  relation and a conservation law. Both of these results can be
  extended to the noncommutative case, with the difference that they
  result from a cocycle condition on the high-frequency contribution
  to the Poisson structure, not from the field equations.}

\thispagestyle{empty}

\vskip10pt
 \newpage
\tableofcontents
\thispagestyle{empty} \newpage

\setlength{\parskip}{5pt plus2pt minus2pt}


\initiate
\section{Introduction and motivation}                   \label{Iam}

Our purpose here is to show that a gravitational field is intimately
associated with a lack of commutativity of the local coordinates of
the space-time structure on which the field is defined.  The
gravitational field can be described by a moving frame with local
components $e^\mu_\alpha$; the lack of commutativity by a commutator
$J^{\mu\nu}$.  To a coordinate $x^\mu$ one associates a conjugate
momentum $p_\alpha$ and to this couple a commutator
$[p_\alpha,x^\mu]$.  The relation we are seeking then can be
succinctly written as the identity
\begin{equation}
[p_\alpha, x^\mu] = e^\mu_\alpha.                 \label{pxe}
\end{equation}
After some preliminary mathematics we shall be in a position to state
the relation in more detail.

Let $\mu$ be a typical `large' source mass with `Schwarzschild radius'
$G_N\mu$. We have two length scales, determined by respectively the
square $G_N\hbar$ of the Planck length and by $\kbar$. The
gravitational field is weak if the dimensionless parameter
$\epsilon_{GF} = G_N \hbar \mu^2$ is small; the space-time is almost
commutative if the dimensionless parameter $\epsilon = \kbar \mu^2$ is
small. These two parameters are not necessarily related but we shall
here assume that they are of the same order of magnitude,
\begin{equation}
\epsilon_{GF} \simeq \epsilon.             \label{ekm}
\end{equation}
If noncommutativity is not directly related to gravity then it makes
sense to speak of ordinary gravity as the limit $\kbar \to 0$ with
$G_N\mu$ nonvanishing. On the other hand if noncommutativity and
gravity are directly related then both should vanish with $\kbar$. We
wish here to consider an expansion in the parameter $\epsilon$, which
we have seen is a measure of the relative dimension of a typical
`space-time cell' compared with the Planck length of a typical
quantity of gravitational energy.  Our motivation for considering
noncommutative geometry as an `avatar' of gravity is the belief that
it sheds light on the role~\cite{Pau56} of the gravitational field as
the universal regulator of ultra-violet divergences.  We give a brief
review of the approach we use to study gravitational fields on
Lorentz-signature manifolds but only in so far as it is necessary. A
general description can be found elsewhere~\cite{Mad00c} as can a
simple explicit solution~\cite{BurMad05b}.

We introduce a set $J^{\mu\nu}$ of elements of an associative algebra
$\c{A}$ (`noncommutative space' or `fuzzy space') defined by
commutation relations
\begin{equation}
[x^\mu, x^\nu] = i\kbar J^{\mu\nu}(x^\sigma) .      \label{xx}
\end{equation}
The constant $\kbar$ is a square of a real number which defines the
length scale on which the effects of noncommutativity become
important.  The $J^{\mu\nu}$ are restricted by Jacobi identities; we
shall see below that there are two other requirements which also
restrict them.

We suppose the differential calculus over $\c{A}$ to be defined by a
frame, a set of 1-forms $\theta^\alpha$ which commute with the
elements of the algebra. We assume the derivations dual to these forms
to be inner, given by momenta $p_\alpha$ as in ordinary quantum
mechanics
\begin{equation}
e_\alpha = \ad p_\alpha.
\end{equation} 
Recall that here the momenta $p_\alpha$ include a factor
$(i\kbar)^{-1}$.  The momenta stand in duality to the position
operators by the relation~(\ref{pxe}).  However, now consistency
relations in the algebra restrict $\theta^\alpha$ and $J^{\mu\nu}$.
Most important thereof is the Leibniz rule which defines differential
relations
\begin{equation}
i\kbar [p_\alpha, J^{\mu\nu}] = [x^{[\mu},[p_\alpha,x^{\nu]}]] 
=  [x^{[\mu}, e^{\nu]}_\alpha]                      \label{Jg}
\end{equation}
between the $J^{\mu\nu}$ on the left and the frame components 
$e_\alpha^\mu$ on the right.

Now we can state the relation (\ref{pxe}) between noncommutativity and
gravity more precisely. The right-hand side of this identity defines
the gravitational field. The left-hand side must obey Jacobi
identities. These identities yield relations between quantum mechanics
in the given curved space-time and the noncommutative structure of the
algebra. The three aspects of reality then, the curvature of
space-time, quantum mechanics and the noncommutative structure are
intimately connected. We shall consider here an even more exotic
possibility that the field equations of general relativity are encoded
also in the structure of the algebra so that the relation between
general relativity and quantum mechanics can be understood by the
relation which each of these theories has with noncommutative
geometry.

We resume the
various possibilities in a diagram, starting with a classical
\nobreak{metric}~$\t{g}_{\mu\nu}$.
\begin{equation}
\begin{array}{ccccccc}
& &\t{g}_{\mu\nu} &\longrightarrow &\t{\theta}^\alpha
&\longleftrightarrow &\t{\Lambda}^\alpha_\mu
\\[15pt]
& & & &\downarrow & &
\\[15pt]
& &g_{\mu\nu} &\longleftarrow &\theta^\alpha
&\longleftrightarrow &\Omega(\c{A})
\\[15pt]
& & & &\downarrow & &
\\[15pt]
& & & &J^{\mu\nu} &\longrightarrow &\c{A}
\end{array}
\end{equation}
The most important flow of information is from the classical metric
$\t{g}_{\mu\nu}$ to the commutator $J^{\mu\nu}$, defined in three
steps.  The first step is to associate to the metric a moving frame
$\t{\theta}^\alpha$, which can be written in the form
$\t{\theta}^\alpha = \t{\theta}^\alpha_\mu dx^\mu$.  The frame is then
`quantized' according to the ordinary rules of quantum mechanics; the
dual derivations $\t{e}_\alpha$ are replaced by inner derivations
$e_\alpha = \ad p_\alpha$ of a noncommutative algebra. The commutation
relations are defined by the $J^{\mu\nu}$, obtained from the
$\theta^\alpha$ by solving a differential equation.  If the space is
flat and the frame is the canonical flat frame then the right-hand
side of~(\ref{Jg}) vanishes and it is possible to consistently choose
$J^{\mu\nu}$ to be constant; the map~(\ref{map}) is not single valued
since any constant $J^{\mu\nu}$ has flat space as inverse image.  If
the noncommutative structure is defined by a twist $S$ then the latter
can be used to define a twisted derivation of the algebra. In many
cases it can be shown to be equivalent to a frame derivation which
satisfies the ordinary Leibniz rule.

There are in fact three enmeshed problems to consider. The first is
the map
 \begin{equation}
J^{\mu\nu} \longrightarrow   \mbox{Curv} (\theta^\alpha) \label{map}
\end{equation}
from the Poisson structure $J^{\mu\nu}$ to the curvature of a frame.
It allows us to express the Einstein tensor in terms of
$J^{\mu\nu}$.  The interest at the moment of this point is limited by
the fact that we have no `equations of motion' for $J^{\mu\nu}$.

The second problem is the mode decomposition of the image metric. We
shall see that in the linear approximation there are three modes in
all, which we shall consider as the two dynamical modes of a spin-2
particle plus a scalar mode. They need however not all be present: the
graviton will be polarized by certain background noncommutative
`lattice' structure.  This leads to the problem of the propagation of
the modes in the `lattice'. The covariant WKB
approximation~\cite{Isa68a,Isa68b,Mad73} applied to the Einstein
equations is an elegant method to study the propagation of
gravitational waves in a given background. We shall mimic it here as
far as possible.

The third problem we wish to aboard here is that of the existence and
definition of an energy-momentum for the Poisson structure and of an
eventual contribution of this energy-momentum to the gravitational
field equations. In the formulation which we are considering the
Einstein tensor is determined by integrability conditions for the
underlying associative-algebra structure, which suggests the
possibility of interpreting the commutative limit of this tensor as
the energy-momentum of the symplectic structure. We have no action for
the metric; the field equations, we claim, are integrability
conditions for the differential calculus. One of these conditions is
the cocycle condition~(\ref{cocon}) given below, which is similar in
structure to the condition that the Ricci tensor vanish; the two are
however not equivalent.

This article is a natural sequel of a previous
one~\cite{BurGraMadZou06a}.  The basic idea was also partially
anticipated in a more specialized treatment~\cite{Mad97c} of
asymptotically-flat space-times as well as in a string-theoretical
reduction~\cite{Sch99,AleRecSch00} of noncommutative geometry to a
supplementary 2-form, the Kalb-Ramond $B$-field, which appears within
the context of super-string theory. There is also a definite overlap
with an interesting recent interpretation~\cite{Ste07} of the
map~(\ref{map}) as a redefinition of the gravitational field in terms
of noncommutative electromagnetism. We shall return to this
interpretation in Section~\ref{covcoord}.

The paper is organized as follows.  In Section~2 we derive the
map~(\ref{map}). In Section~\ref{wkb} we discuss the mode
decomposition using the WKB formalism. In Section~\ref{vtjam} finally
we propose a definition of the energy-momentum of the limiting Poisson
structure.  We use letters $\lambda,\mu,\nu, \rho$ to denote the
coordinate indices and $\alpha, \beta,\gamma, \zeta$ for the frame
indices. In the example we use the frame metric $(-1,1,1,1)$.

\initiate
\section{The Correspondence}

To fix the notation we give briefly some elements of the
noncommutative frame formalism. We refer to the
literature~\cite{Mad00c,BurGraMadZou06a} for further details.

\subsection{Preliminary formalism}

We start with a noncommutative $*$-algebra $\c{A}$ generated by four
hermitian elements $x^\mu$ which satisfy the commutation
relations~(\ref{xx}).  Assume that over $\c{A}$ is a differential
calculus which is such~\cite{Mad00c} that the module of 1-forms is
free and possesses a preferred frame $\theta^\alpha$ which commutes,
\begin{equation}
[x^\mu, \theta^\alpha] = 0                             \label{mod}
\end{equation}
with the algebra. The space one obtains in the commutative limit is
therefore parallelizable with a global moving frame
$\t{\theta}^\alpha$ defined to be the commutative limit of
$\theta^\alpha$.  We can write the differential
\begin{equation}
dx^\mu = e_\alpha^\mu \theta^\alpha, \qquad 
e_\alpha^\mu = e_\alpha x^\mu.
\end{equation}
The differential calculus is defined as the largest one consistent
with the module structure of the 1-forms so constructed.  The algebra
is defined by a product which is restricted by the matrix of elements
$J^{\mu\nu}$; the metric is defined by the matrix of elements
$e_\alpha^\mu$.  Consistency requirements, essentially determined by
Leibniz rules, impose relations between these two matrices which in
simple situations allow us to find a one-to-one correspondence between
the structure of the algebra and the metric.  The input of which we
shall make the most use is the Leibniz rule~(\ref{Jg}) which can also
be written as relation between 1-forms
\begin{equation}
i\kbar d J^{\mu\nu} = [dx^\mu, x^\nu] + [x^\mu, dx^\nu].   \label{lr}
\end{equation}
One can see here a differential equation for $J^{\mu\nu}$ in terms of
$e^\mu_\alpha$. In important special cases the equation reduces to a
simple differential equation of one variable.

In addition, we must insure that the differential is well defined. A
necessary condition is that $d[x^\mu, \theta^\alpha] = 0$, from which
it follows that the momenta $p_\alpha$ must satisfy quadratic
relation~\cite{Mad00c}.  On the other hand, from~(\ref{mod}) it
follows that
\begin{equation}
d[x^{\mu}, \theta^{\alpha}] =
[dx^{\mu}, \theta^{\alpha}] + [x^\mu, d\theta^{\alpha}] =
e^\mu_\beta [\theta^{\beta}, \theta^{\alpha}] -
\tfrac 12 [x^\mu, C^\alpha{}_{\beta\gamma}] \theta^{\beta}\theta^{\gamma},
\end{equation}
where we  have  introduced the Ricci rotation coefficients
\begin{equation}
d\theta^\alpha =-\frac 12 C^\alpha{}_{\beta\gamma} \theta^\beta \theta^\gamma .
\end{equation} 
Therefore we find that multiplication of 1-forms must satisfy
\begin{equation}
[\theta^{\alpha}, \theta^{\beta}] =
\tfrac 12 \theta^\beta_\mu [x^\mu, C^\alpha{}_{\gamma\delta}]
\theta^{\gamma}\theta^{\delta}.                               \label{bffp}
\end{equation}
Using the consistency conditions we obtain that
\begin{equation}
\theta^{[\beta}_\mu
[x^\mu, C^{\alpha]}{}_{\gamma\delta}] = 0 ,                 \label{xC}
\end{equation}
and also that the expression $
\theta^{(\alpha}_\mu [x^\mu, C^{\beta)}{}_{\gamma\delta}] $ must be central.

The metric is defined by the map
\begin{equation}
g(\theta^\alpha\otimes\theta^\beta) = g^{\alpha\beta}.
\end{equation}
The bilinearity of the metric implies that $g^{\alpha\beta}$ are
complex numbers.  We choose the frame to be orthonormal in the
commutative limit; we can write therefore
\begin{equation}
g^{\alpha\beta} = \eta^{\alpha\beta} - \ep h^{\alpha\beta}.  \label{hh}
\end{equation}

We introduce also
\begin{equation}
g^{\mu\nu} = g(dx^\mu \otimes dx^\nu)
= e^\mu_\alpha e^\nu_\beta g^{\alpha\beta}.
\end{equation}
We write $g^{\mu\nu}$ as a sum
\begin{equation}
g^{\mu\nu} = g_+^{\mu\nu} + g_-^{\mu\nu}
\end{equation}
of symmetric and antisymmetric parts. To lowest order in the
noncommutativity in general we have $h^{\alpha\beta} = -h^{\beta\alpha}$ so
we find that
\begin{equation}
g_+^{\mu\nu}
= \tfrac 12 \eta^{\alpha\beta} [e^{\mu}_\alpha, e^{\nu}_\beta]_+
- \tfrac 12 \ep h^{\alpha\beta} [e^{\mu}_\alpha, e^{\nu}_\beta]
\end{equation}
and
\begin{equation}
g_-^{\mu\nu}
= \tfrac 12 \eta^{\alpha\beta} [e^{\mu}_\alpha, e^{\nu}_\beta]
- \tfrac 12  \ep h^{\alpha\beta} [e^{\mu}_\alpha, e^{\nu}_\beta]_+.
\end{equation}
We shall restrict our considerations in Section~\ref{gf} to
first-order perturbations of flat space. We set
\begin{equation}
g^{\mu\nu} = \eta^{\mu\nu} - \epr g_1^{\mu\nu}, \qquad
e^\mu_\alpha = \delta^\mu_\alpha + \epr \Lambda^\mu_\alpha.
\end{equation}
We have then the relations
\begin{equation}
 g_1^{\mu\nu} =
- \eta^{\alpha\beta} \Lambda^{(\mu}_\alpha \delta^{\nu)}_\beta
= - \Lambda^{(\mu\nu)}.
\end{equation}

\subsection{The quasi-commutative approximation}         \label{g2a}

To lowest order in $\epr$ the partial derivatives are well defined and
the approximation, which we shall refer to as the quasi-commutative,
\begin{equation}
[x^\lambda, f] =
i\kbar J^{\lambda\sigma} \p_\sigma f                \label{der}
\end{equation}
is valid.  The Leibniz rule and the Jacobi identity can be written in
this approximation as
\begin{eqnarray}
&& e_\alpha J^{\mu\nu}
= \partial_\sigma e^{[\mu}_\alpha J^{\sigma\nu ]}  ,         \label{L*}
\\[6pt]&&
\epsilon_{\kappa\lambda\mu\nu} J^{\gamma\lambda}e_\gamma J^{\mu\nu}=0. 
                                                        \label{J*}
\end{eqnarray}
We shall refer to these equations including their integrability
conditions as the Jacobi equations.

Written in frame components the Jacobi equations become
\begin{eqnarray}
&&
e_\gamma J^{\alpha\beta}
- C^{[\alpha}{}_{\gamma\delta} J^{\beta]\delta} = 0,      \label{cl}
\\[6pt]&&
\epsilon_{\alpha\beta\gamma\delta}J^{\gamma\eta}
(e_\eta J^{\alpha\beta}
+ C^\alpha{}_{\eta\zeta} J^{\beta\zeta}) = 0.              \label{mb***}
\end{eqnarray}
We have used here the expression for the rotation coefficients, 
valid also in the quasi-commutative approximation:
\begin{equation}
C^\alpha{}_{\beta\gamma} 
= \theta^\alpha_\mu e_{[\beta}e^\mu_{\gamma]}
=-e^\nu_\beta e^\mu_\gamma\partial_{[\nu}\theta^\alpha_{\mu]}.\label{rc}
\end{equation}
Inserting (\ref{cl}) into (\ref{mb***}) one finds the relation
\begin{equation}
\epsilon_{\alpha\beta\gamma\delta}
J^{\alpha\zeta}J^{\beta\eta}C^{\gamma}{}_{\eta\zeta} = 0.   \label{b**}
\end{equation}
One can solve (\ref{cl}) for the rotation coefficients. One obtains
\begin{equation}
J^{\gamma\eta}e_\eta J^{\alpha\beta} =
J^{\alpha\eta}J^{\beta\zeta}C^\gamma{}_{\eta\zeta},
\end{equation}
or, provided $J^{-1}$ exists, 
\begin{equation}
C^\alpha{}_{\beta\gamma}
 = J^{\alpha\eta} e_\eta J^{-1}_{\beta\gamma}.            \label{mbb}
\end{equation}
This can be rewritten as
\begin{equation}
C^\alpha{}_{\beta\gamma}
 = J^{\alpha\delta} e_\delta J^{\zeta\eta}
J^{-1}_{\zeta\beta} J^{-1}_{\eta\gamma}.                   \label{mbbj}
\end{equation}
From general
 considerations also follows that the rotation coefficients must 
satisfy the gauge condition
\begin{equation}
e_\alpha C^{\alpha}{}_{\beta\gamma} =  0  .                 \label{gauge-co}
\end{equation}

Anticipating a notation from Section~\ref{gf} we introduce
\begin{equation}
\h{C}_{\alpha\beta\gamma} =
J^{-1}_{\alpha\delta} C^\delta{}_{\beta\gamma}.
\end{equation}
We find then that
\begin{equation}
\h{C}_{\alpha\beta\gamma} =  e_\alpha J^{-1}_{\beta\gamma}
\end{equation}
and that
\begin{equation}
\h{C}_{\alpha\beta\gamma} + \h{C}_{\beta\gamma\alpha}
+ \h{C}_{\gamma\alpha\beta} = 0,                \label{dR}
\end{equation}
an equation which is in fact obviously the same as~(\ref{b**}) and
which we can write as a de~Rham cocyle condition
\begin{equation}
d J^{-1} = 0, \qquad J^{-1} 
= \frac 12 J^{-1}_{\alpha\beta} \theta^\alpha\theta^\beta. \label{cocon}
\end{equation}
One can write Equation~(\ref{mbb}) as
\begin{equation}
e_\alpha J^{-1}_{\beta\gamma} =
J^{-1}_{\alpha\delta}C^\delta{}_{\beta\gamma} .
\end{equation}

It follows that in the quasi-classical approximation the linear
connection and therefore the curvature can be directly expressed in
terms of the commutation relations.  This is the content of the map
(\ref{map}).  Using the expression
\begin{equation}
\omega_{\alpha\beta\gamma} 
=\tfrac 12 (C_{\alpha\beta\gamma} 
- C_{\beta\gamma\alpha} + C_{\gamma\alpha\beta})
\end{equation}
for the Ricci curvature tensor for example we obtain 
\begin{eqnarray}
&&
2R_{\beta\zeta}
= J_{(\zeta\delta }e^\alpha e^\delta J^{-1}_{\beta)\alpha }
+ J^{\alpha\delta }e_{(\zeta}e_\delta J^{-1}_{\alpha\beta)}
\nonumber\\[8pt]&&\phantom{2R_{\alpha\zeta}}
- J_{(\zeta }{}^\eta e^\alpha J^{-1}_{\eta\gamma}
J^{\gamma\delta }e_\delta J^{-1}_{\beta)\alpha}
+ J^{\alpha\delta }e_\delta J^{-1}_{\eta\beta}
J^{\eta\gamma }e_\gamma J^{-1}_{\alpha\zeta}
\nonumber\\[8pt]&&\phantom{2R_{\alpha\zeta}}
+ J_{\eta\delta }e^\delta J^{-1}_{\beta\alpha}
J^{\eta\gamma }e_\gamma J^{-1}_{\zeta}{}^\alpha
+ J^{\alpha\eta }e_{(\zeta} J^{-1}_{\eta\gamma}
J^{\gamma\delta }e_\delta J^{-1}_{\beta)\alpha}
\nonumber\\[8pt]&&\phantom{ 2R_{\alpha\zeta}}
-\tfrac 12
J_{\zeta\delta }e^\delta J^{-1 \alpha\eta}
J_{\beta\gamma }e^\gamma J^{-1}_{\alpha\eta}
+ J^{ \alpha\delta}e_\delta J^{-1}_{\alpha\eta}
J_{(\zeta\gamma }e^\gamma J^{-1}_{\beta)}{}^\eta.     \label{R2ndo}
\end{eqnarray}
We have here neglected the ordering on the right-hand side as it gives
the corrections of second-order in $\epr$. To understand better the
relation between the commutator and the curvature in the following
section we shall consider a linearization about a fixed `ground
state'.

\subsection{The weak-field  approximation}  \label{gf}

Assume then that we have a `ground state' consisting of a $J_0$ and a
corresponding image  $\theta_0^\alpha$ of the map 
\begin{equation}
J_0 \rightarrow \mbox{Curv}(\theta_0^\alpha)      \label{map0}
\end{equation}
which we can extend to a region around $J_0$.  We assumed in
Section~\ref{g2a} that the noncommutativity is small and we derived
some relations to first-order in the parameter $\epr$.  We shall now
make an analogous assumption concerning the gravitational field; we
shall assume that $\epsilon_{GF}$ is also small and of the same order
of magnitude. With these two assumptions the Jacobi equations become
relatively easy to solve.

We suppose that the basic unknowns, $J_0$ and $\theta^\alpha_0$  are
constants and that they are perturbed to:
\begin{equation}
J^{\alpha\beta} = J_0^{\alpha\beta} + \epr I^{\alpha\beta}, \qquad 
\theta^\alpha = \theta^\beta_0 (\delta_\beta^\alpha -
 \epr \Lambda_\beta^\alpha).
\end{equation} 
The leading order of the Jacobi system is given by
\begin{eqnarray}
&&
e_\gamma I^{\alpha\beta}
- e_{[\gamma}\Lambda^{[\alpha}_{\delta]} J_0^{\beta]\delta} = 0,        \label{=clp}
\\[6pt]&&
\epsilon_{\alpha\beta\gamma\delta}J_0^{\gamma\eta}
e_\eta I^{\alpha\beta} = 0.                                       \label{mb=p}
\end{eqnarray}
Introducing the notation
\begin{equation}
\h{I}_{\alpha\beta} =
J_0^{-1}{}_{\alpha\gamma} J_0^{-1}{}_{\beta\delta} I^{\gamma\delta}, \qquad
\h{\Lambda}_{\alpha\beta}
= J_0^{-1}{}_{\alpha\gamma}\Lambda^\gamma_\beta,             \label{Hat}
\end{equation}
(\ref{=clp}) can be written as
\begin{equation}
e_\gamma(\h{I} _{\alpha\beta} - \h{\Lambda} _{[\alpha\beta]})
= e_{[\alpha}\h{\Lambda}_{\beta]\gamma}                        \label{mb*2x}
\end{equation}
and~(\ref{mb=p})  as
\begin{equation}
\epsilon^{\alpha\beta\gamma\delta} e_\alpha \h{I}_{\beta\gamma}
= 0                  .                                           \label{c12x}
\end{equation}
We note that $\hI$ is a linear perturbation of $J_0^{-1}$,
\begin{equation}
J_{\alpha\beta}^{-1} = J_{0\alpha\beta}^{-1} + \epr \hI_{\alpha\beta}.
\end{equation}                              
Equations~(\ref{mb*2x}-\ref{c12x}) are the origin of the
particularities of our construction, along with the fact that the
`ground-state' value $J_0^{\mu\nu}$ is an invertible matrix.

We decompose $\h{\Lambda}$ as the sum
\begin{equation}
\h{\Lambda}_{\alpha\beta} 
= \h{\Lambda}^+_{\alpha\beta} + \h{\Lambda}^-_{\alpha\beta}
\end{equation}
of a symmetric and antisymmetric term. Constraint~(\ref{mb*2x}) can be
written then as
\begin{equation}
e_\gamma\h{I} _{\alpha\beta}
- (e_{\alpha}\h{\Lambda}^-_{\beta\gamma}
+ e_{\beta}\h{\Lambda}^-_{\gamma\alpha}
+ e_{\gamma}\h{\Lambda}^-_{\alpha\beta})
= e_{\gamma}\h{\Lambda}^-_{\alpha\beta}
+ e_{[\alpha} \h{\Lambda}^+_{\beta]\gamma}.                \label{*2b3}
\end{equation}
It implies a second cocyle condition. If we multiply by
$\epsilon^{\alpha\beta\gamma\delta}$ we find that
\begin{equation}
\epsilon^{\alpha\beta\gamma\delta} e_\gamma \h{\Lambda}^-_{\alpha\beta}
= 0,
\end{equation}
and Equation~(\ref{*2b3}) simplifies to
\begin{equation}
e_\gamma(\h{I} _{\alpha\beta}- \h{\Lambda}^-_{\alpha\beta})
= e_{[\alpha} \h{\Lambda}^+_{\beta]\gamma}.                 \label{+3}
\end{equation}
This equation has the integrability conditions
\begin{equation}
e_\delta e_{[\alpha} \h{\Lambda}^+_{\beta]\gamma}
- e_\gamma e_{[\alpha} \h{\Lambda}^+_{\beta]\delta}  = 0.      \label{+3a}
\end{equation}
But the left-hand side is the linearized approximation to the
curvature of a (fictitious) metric with components 
$g_{\mu\nu} + \epr\h{\Lambda}^+_{\mu\nu}$. If it vanishes then the
perturbation is a derivative. (We have in fact shown here that a
deformation of a commutator can be always chosen antisymmetric to
first order). For some 1-form $A$ with frame components $A_\alpha$
\begin{equation}
\h{\Lambda}^+_{\beta\gamma} = \tfrac 12 e_{(\beta} A_{\gamma)}.   \label{dL+}
\end{equation}
Equation~(\ref{+3}) becomes therefore
\begin{equation}
e_\alpha (\h{I} - \h{\Lambda}^- - dA)_{\beta\gamma} = 0.    \label{+}
\end{equation}
It follows then that for some 2-form $c$ with constant components
$c_{\beta\gamma}$
\begin{equation}
\h{\Lambda}^- = \h{I} - dA + c, \qquad
\h{\Lambda}_{\alpha\beta} = \h{I}_{\alpha\beta}
+ e_\beta A_\alpha + c_{\alpha\beta}.                           \label{LI}
\end{equation}
The remaining constraints are satisfied identically.

We can also introduce 2-forms
\begin{equation}
\h{I} = \tfrac 12 \h{I}_{\alpha\beta}\theta^\alpha \theta^\beta, \qquad
\h{\Lambda}^- = \tfrac 12 \h{\Lambda}^-_{\alpha\beta}\theta^\alpha \theta^\beta
\end{equation}
and write
\begin{equation}
d\h{I} = 0, \qquad d\h{\Lambda}^- = 0.                  \label{dI}
\end{equation}
The first equation is a particular case of Equation~(\ref{cocon}).
Because of the cocycle condition there must exist a 1-form $C$ such that
\begin{equation}
\h{I}_{\gamma\delta} = e_{[\gamma}C_{\delta]}.              \label{j12}
\end{equation}
We shall see below that the form of $C$ depends partially on the
choice of coordinates.

\subsection{The map}

We can now be more precise about the map~(\ref{map}) in the
linear approximation.  We recall that  the  fluctuations 
 are redundantly parameterized by
$\Lambda^\alpha_\beta$. We can rewrite the map~(\ref{map}) as a map
\begin{equation}
\hI^{\alpha\beta} \longrightarrow \Lambda^\alpha_\beta           \label{wh}
\end{equation}
from $\hI^{\alpha\beta}$ to $\Lambda^\alpha_\beta$. With $c_{\alpha\beta}=0$ we have
\begin{equation}
\Lambda^\alpha_\beta
= J_0^{\alpha\gamma}( \h{I}_{\gamma\beta} 
+ e_\beta A_\gamma).                                    \label{why}
\end{equation}
As a perturbation of the frame 
$e^\mu_\alpha = \delta^\mu_{\beta} + \epr\Lambda^\mu_\alpha$
engenders a perturbation of the metric
\begin{equation}
g^{\mu\nu} =  \eta^{\mu\nu} - \epr g_1{}^{\mu\nu}, \qquad
g_1{}^{\mu\nu} = -\Lambda^{(\mu\nu)} ,
\end{equation}
it follows from~(\ref{why}) that
\begin{equation}
g_1{}_{\alpha\beta} = - J_{0(\alpha}{}^{\gamma}
(\h{I}_{\gamma\beta)} + e_{\beta)} A_\gamma)       .       \label{hI}
\end{equation}
 The correction (\ref{hh})
will appear only in second order.
The frame itself is given by
\begin{equation}
\theta^\alpha = d ( x^\alpha -\epr J_0^{\alpha\gamma}A_\gamma)  
- \epr J_0^{\alpha\gamma} \h{I}_{\gamma\beta} dx^\beta .
\end{equation}
We therefore find  the following expressions
\begin{eqnarray}
&& d\theta^\alpha
= - \epr J_0^{\alpha\gamma}e_\delta \h{I}_{\gamma\beta} dx^\delta dx^\beta 
= \tfrac 12 \epr J_0^{\alpha\delta}e_\delta \hI_{\beta\gamma}dx^\gamma
dx^\beta , 
\\[6pt]&&  
C^\alpha{}_{\beta\gamma }= 
\epr J_0^{\alpha\delta} e_\delta\h{I}_{\beta\gamma},
\\[6pt]&& 
\omega_{\alpha\beta\gamma}
= \tfrac 12 \epr (J_{0[\alpha}{}^\delta e_\delta\h{I}_{\beta\gamma]}
+ J_{0\beta}{}^\delta e_\delta\h{I}_{\alpha\gamma}).  \label{oJ}
\end{eqnarray}
The torsion obviously vanishes.

From~(\ref{oJ}) for the linearized Riemann tensor we obtain, using the
cocycle condition, the expression
\begin{equation}
R_{\alpha\beta\gamma\delta} = \tfrac{1}{2}\epr e^\eta \left(
J_0{}_{\eta[\gamma} e_{\delta]}\h{I}_{\alpha\beta} 
+ J_0{}_{\eta[\alpha} e_{\beta]}\h{I}_{\gamma\delta} 
\right)         .                        \label{vtj4}
\end{equation}
For the  Ricci
curvature we find
\begin{equation}
R_{\beta\gamma } = - \tfrac{1}{2}\epr e^\zeta \left(
J_0{}_{\zeta(\beta} e^\alpha \h{I}_{\gamma)\alpha} 
+ J_0^{\alpha}{}_{\zeta} e_{(\beta} \h{I}_{\gamma)\alpha} \right). 
                                                        \label{vtj5}
\end{equation}
One more contraction yields the expression
\begin{equation}
R = - 2\epr J_0{}^{\zeta\alpha}e_\zeta e^\beta  
\h{I}_{\alpha\beta}                                     \label{vtj6}
\end{equation}
for the Ricci scalar.  Using again the cocycle condition permits us to
write this in the form
\begin{equation}
R =  \epr  \Delta \chi ,           \label{vtj7}
\end{equation}
where the scalar trace component is defined as
\begin{equation}
\chi = J_0^{\alpha\beta}\h{I}_{\alpha\beta}        .      \label{ch}
\end{equation}
The Ricci scalar is a divergence. Classically it vanishes when the
field equations are satisfied.

\subsection{Coordinate invariance}                        \label{ci}

In principle, because we use the frame formalism, covariance is
assured at least to the semi-classical approximation. (We are not sure
exactly what this would mean in general.) However since we base our
construction on the commutator of two generators (coordinates) it is
of interest to show this explicitly.  The frame components
$J_1^{\alpha\beta}$ of the perturbation of the commutator are not
equal to the perturbation $I^{\alpha\beta}$ of the frame components
but the two are related in a simple way. We have to first order the
relation
\begin{equation}
J^{\alpha\beta} = \theta^\alpha_\mu \theta^\beta_\nu J^{\mu\nu}
\end{equation}
from which we conclude that
\begin{eqnarray}
&&
J_1^{\alpha\beta} = I^{\alpha\beta}
- J_0^{[\alpha\gamma}\Lambda^{\beta]}_\gamma 
= I^{\alpha\beta} - J_0^{[\alpha\gamma}J_0^{\beta]\delta} \h{\Lambda}_{\delta\gamma}
\nonumber\\[4pt]&&\phantom{J_1^{\alpha\beta}}
= - (I^{\alpha\beta} 
+ J_0^{\alpha\gamma}J_0^{\beta\delta}
e_{[\gamma} A_{\delta]} )
\nonumber\\[4pt]&&\phantom{J_1^{\alpha\beta}}
= - J_0^{\alpha\gamma}J_0^{\beta\delta}
e_{[\gamma} (C+A)_{\delta]}.                                        \label{j1}
\end{eqnarray}
The second line is obtained using the solution~(\ref{why}) and the last
uses the expression~(\ref{j12}).

If we consider a first-order coordinate transformation of the form
\begin{equation}
x^{\prime\mu} = x^\mu + \epr B^\mu
\end{equation}
we conclude that under this transformation the components of $A$ transform as
\begin{equation}
A^\prime_\alpha = A_\alpha + J^{-1}_{0\alpha\beta} B^\beta.
\end{equation}
From the last line of the sequence of identities~(\ref{j1}) we see
that we can choose the coordinates to set $J_1^\prime=0$. 
This is the Darboux theorem. We can also choose coordinates to set $A^\prime =0$.
We cannot however set $I=0$, fortunately since this would
entail that the curvature vanish.

\subsection{Covariant coordinates}      \label{covcoord}

There is a special case of particular interest, that in which the
mixed components $J_0^{\mu\alpha}$ are constant and the matrix they
form is invertible. We write then
\begin{equation}
x^\mu = J_0^{\mu\alpha} D_\alpha, \qquad 
D_\alpha = p_\alpha + \cA_\alpha.                          \label{xpA}
\end{equation}
The interest in this decomposition resides in the properties of
the 1-forms $\cA = \cA_\alpha \theta^\alpha$ and 
$\theta = - p_\alpha \theta^\alpha$ considered as gauge potentials. 
Let $\c{U} \subset \c{A}$ be the group of unitary elements of the
algebra and define for arbitrary $\cA$ and $g\in \c{U}$
\begin{equation}
\cA^\prime = g^{-1} \cA g + g^{-1} dg.
\end{equation}
Since
\begin{equation}
dg = e_\alpha g \theta^\alpha = - [\theta, g]
\end{equation}
in the particular case with $\cA = \theta$ we have 
$\theta^\prime = \theta$. We conclude that, being the difference between
two gauge potentials, the generators $x^\mu$ transform as adjoint
representations of $\c{U}$:
\begin{equation}
x^{\prime\mu} = g^{-1} x^\mu g.
\end{equation}
This decomposition was introduced~\cite{DubKerMad89b} in the 
particular case of a matrix algebra to describe the shift caused by
spontaneous symmetry breaking and the generators were 
called~\cite{MadSchSchWes00a,MadSchSchWes00b} covariant because of
their transformation properties.

From~(\ref{xpA}) we deduce that
\begin{equation}
J^{\mu\nu} = J_0^{\mu\alpha}J_0^{\nu\beta} F_{\alpha\beta}
\end{equation}
with
\begin{equation}
F = \frac 12  F_{\alpha\beta} \theta^\alpha\theta^\beta = d\cA  + \cA^2 .
\end{equation}
It would seem natural in this case at least to identify~\cite{Ste07}
noncommutativity as noncommutative electromagnetism and consider the
action 
\begin{equation}
S = \frac 12 \, \tr J_{\mu\nu} J^{\mu\nu}
\end{equation}
as the action for commutator. We recall that noncommutative
electromagnetism can contain Yang-Mills components with gauge group
$U_n$ for arbitrary $n$. It suffices that the algebra $\c{A}$ contain
a factor $M_n$.

\initiate
\section{The WKB approximation}                \label{wkb}

In the commutative case the WKB dispersion relations followed from the
field equations. In order to introduce the WKB approximation in 
noncommutative case,
we suppose that the algebra $\c{A}$ is a tensor product
\begin{equation}
\c{A} = \c{A}_0 \otimes \c{A}_\omega
\end{equation}
of a `slowly-varying' factor $\c{A}_0$ in which all amplitudes lie and
a `rapidly-varying' phase factor which is of order-of-magnitude $\epr$
so that only functions linear in this factor can appear. By
`slowly-varying' we mean an element $f$ with a classical limit $\t{f}$
such that $\p_\alpha \t{f} \lesssim \mu \t{f}$.  The generic element
$f$ of the algebra is of the form then
\begin{equation}
f = f_0 + \epr \bar f e^{i\omega\phi}
\end{equation} 
where $f_0$ and $\bar f$ belong to $\c{A}_0$.  Because of the condition
on $\epsilon$ the factor order does not matter and these elements form
an algebra.  The frequency parameter $\omega$ is so chosen that for an
element $f$ of $\c{A}_0$ the estimate
\begin{equation}
[\phi, f] \simeq \kbar \mu
\end{equation}
holds. The commutator $[f, e^{i\omega\phi}]$ is thus of order of
magnitude
\begin{equation}
[f, e^{i\omega\phi}] \simeq \kbar \mu\omega .
\end{equation}
The wave vector
\begin{equation}
\xi_\alpha = e_\alpha \phi                 \label{ksi}
\end{equation} 
is normal to the surfaces of constant phase.  We shall require also
that the energy of the wave be such that it contribute not as source
to the background field. This inequality can be written as
\begin{equation}
\epsilon \omega^2 \ll \mu^2.                            \label{we}
\end{equation}
It assures us also that to the approximation we are considering we
need not pay attention to the order of the factors in the
perturbation.  We have in fact partially solved the system of
equations without further approximation. One purpose of the following
analysis is to verify that all constraints have been satisfied.  We
first recall the results one obtains in the classical case.

\subsection{The commutative case}

Classically, the vacuum equations are given by the condition that the
Einstein tensor vanish. In the WKB approximation this yields in fact two
equations. The leading order term, proportional to $\omega^2$ is
a dispersion relation (`E' for Einstein)
\begin{equation}
G_{\alpha\beta} =   \tfrac 12  \epr  (i\omega)^2
\mbox{Disp}_E{}_{\alpha\beta} = 0
\end{equation}
with
\begin{equation}
\mbox{Disp}_E{}_{\alpha\beta} = - \xi^2 \psi_{\alpha\beta}
+ \xi^\gamma \psi_{\gamma(\alpha} \xi_{\beta)} -
\xi^\gamma\xi^\delta\psi_{\gamma\delta}\eta_{\alpha\beta},
\end{equation}
and
\begin{equation}
\psi_{\alpha\beta}  = g_1{}_{\alpha\beta}
- \tfrac 12 g^T \eta_{\alpha\beta}, \qquad 
g^T = g_{1\alpha}{}^\alpha.                             \label{EVac1}
\end{equation}
If $\xi^2 = 0$ then it follows that
\begin{equation}
\xi^\gamma \psi_{\gamma\beta} = 0.                            \label{EVac1a}
\end{equation}
The second term in the expansion in frequency, the one proportional to
$\omega$, yields a conservation law
\begin{equation}
G_{\alpha\beta}
= \tfrac 12  \epr  (i\omega) \mbox{Cons}_E{}_{\alpha\beta} = 0 \label{EVac2}
\end{equation}
with
\begin{equation}
\mbox{Cons}_E{}_{\alpha\beta}
= 2 \xi^\gamma e_\gamma \psi_{\alpha\beta}
+  e_\gamma\xi^\gamma \psi_{\alpha\beta}.
\end{equation}
This second equation can be interpreted~\cite{Isa68a,Isa68b,Mad73} as a
conservation of graviton number {\it in vacuo}. One easily sees that
from it follows
\begin{equation}
\mbox{Cons}_E{}_{\beta\gamma} \psi^{\beta\gamma} =
e_\alpha (\psi_{\beta\gamma} \psi^{\beta\gamma} \xi^\alpha) = 0 .
\end{equation}
With the Jacobi equations there is a similar doubling.

\subsection{The quasi-commutative case}

In the WKB approximation the perturbations 
$\Lambda^\alpha_\beta$ and $I^{\alpha\beta}$ are of the form
\begin{equation}
\Lambda^\alpha_\beta 
= {\a{\Lambda}}^\alpha_\beta e^{i\omega\phi}, \qquad
I^{\alpha\beta} = \a{I}^{\alpha\beta} e^{i\omega\phi},
\end{equation}
where $\a{\Lambda}^\alpha_\beta$ and $\a{I}^{\alpha\beta}$ belong to
$\c{A}_0$.  Therefore we have also
\begin{equation}
g_1^{\mu\nu} = \a{g}^{\mu\nu}e^{i\omega\phi} .
\end{equation}
Using $\xi_\alpha = e_\alpha \phi$ and $  
\eta^\alpha = J_0^{\alpha\beta} \xi_\beta    $ we have 
\begin{eqnarray}
&&
e_{\alpha} I_{\beta\gamma}
= (i\omega\xi_\alpha \a{I}_{\beta\gamma}
+ e_{\alpha}\a{I}_{\beta\gamma}) e^{i\omega\phi} ,
\\[4pt]&&
e_{\alpha} \Lambda_{\beta\gamma}
= (i\omega\xi_\alpha \a{\Lambda}_{\beta\gamma}
+ e_{\alpha}\a{\Lambda}_{\beta\gamma}) e^{i\omega\phi}.
\end{eqnarray}

The cocycle condition replaces Einstein's equation to a certain
extent. In the WKB approximation it becomes
\begin{equation}
\xi_\alpha \h{I}_{\beta\gamma}
+ \xi_\beta \h{I}_{\gamma\alpha}
+ \xi_\gamma \h{I}_{\alpha\beta} = 0.                    \label{cyc}
\end{equation}
We multiply this equation by $\xi^\alpha$ and obtain
\begin{equation}
\xi^2 \h{I}_{\beta\gamma} 
+ \xi_{[\beta} \h{I}_{\gamma]\alpha}\xi^\alpha = 0.      \label{cyc2}
\end{equation}
If $\xi^2 \neq 0$ then we conclude that 
\begin{equation}
\h{I}_{\beta\gamma} 
= - \xi^{-2}\xi_{[\beta} \h{I}_{\gamma]\alpha}\xi^\alpha. 
\end{equation}
This is no restriction; it defines simply $C_\alpha$ by
\begin{equation}
i\omega C_{\alpha} = - \xi^{-2} \h{I}_{\alpha\beta}\xi^\beta. 
\end{equation}
If $\xi^2 = 0$ then we conclude that 
\begin{equation}
\xi_{[\beta} \h{I}_{\gamma]\alpha}\xi^\alpha = 0. 
\end{equation}
This is a small restriction; the $\xi_\alpha$ must be a Petrov vector
of $\h{I}$. We shall improve on this in a particular case in
Section~\ref{vtjam}.  In terms of the scalar $\chi$ we obtain the
relation
\begin{equation}
\h{I}_{\alpha\beta}\eta^\beta 
= -\tfrac 12 \chi \xi_\alpha.                           \label{ec}
\end{equation}

Using the definition of $\eta$ we find in the WKB approximation to
first order 
\begin{eqnarray}
&&\omega_{\alpha\beta\gamma}
= \tfrac 12 i\omega\epr \left(\eta_{[\alpha} \h{I}_{\beta\gamma]}
+\eta_{\beta} \h{I}_{\alpha\gamma}\right),                 
\label{oJ1}\\[6pt]&& 
R_{\alpha\beta\gamma\delta} =- \tfrac 12\epr (i\omega)^2 
\left(\eta_{[\gamma}\xi_{\delta]} \hI_{\alpha\beta}
- \eta_{[\alpha}\xi_{\beta]} \h{I}_{\gamma\delta} \right) ,
\label{R}\\[8pt]&&
R_{\beta\gamma} =- \tfrac 12\epr (i\omega)^2 
\left(\xi_{(\beta} \eta^\alpha 
- \xi^{\alpha} \eta_{(\beta}\right) \h{I}_{\gamma)\alpha} ,
\label{cts2b}\\[8pt]&&
 R =  \epr(i\omega)^2\chi\xi^2 .              \label{cts}
\end{eqnarray}
In average, the linear-order expressions  vanish. We can calculate to second order if we average over several
wavelengths. We use the approximations
\begin{equation}
\vev{\h{I}^{\alpha\beta}} = 0, \qquad
 \vev{\h{I}^{\alpha\beta} \h{I}^{\gamma\delta}}
= \frac 12 \h{\a{I}}^{\alpha\beta} \h{\a{I}}^{\gamma\delta} .
\end{equation}
Also as $e_\delta J^{-1}_{\beta\gamma}
= -J^{-1}_{\beta\eta}e_\delta J^{\eta\zeta}\, J^{-1}_{\zeta\gamma}$ 
we can write 
$e_\delta J^{-1}_{\beta\gamma} = \epr e_\delta \h{I}_{\beta\gamma}$.
Therefore we find  expanding~(\ref{R2ndo}) to second order
the expression
\begin{equation}
\vev{R_{\beta\gamma}}
= \tfrac 12\epr^2(i\omega)^2 
\Big( \a{\chi} \xi^\alpha\eta_{(\gamma }\h{\a{I}}_{\beta )\alpha}
+ \tfrac 34  \a{\chi}^2 \xi_\beta\xi_\gamma
+\eta^2\h{\a{I}}_{\eta\beta}\h{\a{I}}^\eta{}_\gamma
-\tfrac 12 \eta_\beta \eta_\gamma
\h{\a{I}}_{\alpha\eta}\h{\a{I}}^{\alpha\eta} \Big) \label{Ric}
\end{equation}
for the Ricci tensor and the expression
\begin{equation}
\vev{R} = \tfrac 18 \epr^2 (i\omega)^2  
(2 \eta^2 \h{\a{I}}_{\alpha\beta}\h{\a{I}}^{\alpha\beta} 
+ 7\a{\chi}^2\xi^2)
\end{equation}
for the Ricci scalar. We shall return to these formulae in
Section~\ref{vtjam}.

\subsection{The noncommutative lattice}                    \label{ncl}

As a lattice, the background noncommutativity is of considerable
complexity, the contrary of a simple cubic lattice. It is in general
non-periodic but in the WKB approximation we can assume periodicity
since at the scale of the frequency $J_0$ is a constant $4\times4$
matrix. It is difficult to obtain general expressions for the modes and
their dispersion relations; however, it is interesting to analyse them in more
detail by considering a specific example. We take an arbitrary
perturbation $\hI_{\alpha\beta}$ with the wave vector $\xi_\alpha$
normalized so that $\xi_0= -1$,
\begin{equation}
\h{\a{I}}_{\alpha\beta} =
\left(\begin{array}{cccc}
   0  &  b_3 & -b_2 & e_1 \\[4pt]
 -b_3 &  0   &  b_1 & e_2 \\[4pt]
  b_2 & -b_1 &  0   & e_3 \\[4pt]
 -e_1 & -e_2 & -e_3 &  0
\end{array}\right) .
\end{equation}
One easily sees that the cocyle condition is equivalent to the
constraint $\vec b = - \vec\xi \times \vec e$ which is the part of the
field equations for a plane wave, the Bianchi equations.  
Suppose that $\xi$ is null and oriented along the $z$-axis,
$\xi_\alpha = (0,0,1,-1)$. If $\hI$ satisfies the cocycle condition we have 
\begin{equation}
\h{\a{I}}_{\alpha\beta} =
\left(\begin{array}{cccc}
   0  &  0   & -e_1 & e_1 \\[4pt]
   0  &  0   & -e_2 & e_2 \\[4pt]
  e_1 & e_2  &  0   & e_3 \\[4pt]
 -e_1 & -e_2 & -e_3 &  0
\end{array}\right)  .
\end{equation}
The perturbation $\hI$ is of Petrov-type $N$ if $\vec\xi\cdot\vec e=e_3=0\,$; 
this would be the second half of the Maxwell field equations.  In
this case, for arbitrary background noncommutativity given by
\begin{equation}
J_{0\alpha\beta}= 
\left(\begin{array}{cccc}
  0  &  B_3 & -B_2 & E_1 \\[4pt]
-B_3 &  0   &  B_1 & E_2 \\[4pt]
 B_2 & -B_1 &  0   & E_3 \\[4pt]
 - E_1 &  -E_2 &  -E_3 &  0
\end{array}\right)
\end{equation}
we can write the amplitude of the metric perturbation in the form
\begin{equation}
\a{g}_{1\alpha\beta} 
= - J_{0(\alpha}{}^{\gamma}\h{\a{I}}_{\gamma\beta)} =
\left(\begin{array}{cc}
P_{11}&P_{12}\\[4pt]P^T_{12}&P_{22}
\end{array}\right).
\end{equation}
It is easy to check that by a change of coordinates we can set 
$P_{12} =0$, $P_{22} =0$. Introducing 
$e_1 = a\cos \gamma$, $e_2 = a\sin\gamma$, 
$B_2+E_1 = A\sin \Gamma$, $B_1-E_2 = A\cos\Gamma$ 
the remaining part $P_{11}$
 can be decomposed 
\begin{equation}
 P_{11} = 
aA\left(\begin{array}{cc}
\sin(\gamma+\Gamma)    &\cos(\gamma+\Gamma)\\[4pt]
\cos(\gamma+\Gamma)    &-\sin(\gamma+\Gamma)
\end{array}\right)
+
aA\left(\begin{array}{cc}
\sin(\gamma-\Gamma)    &         0\\[4pt]
         0             &\sin(\gamma-\Gamma)
\end{array}\right)
\end{equation}
into a trace-free part and a trace. The gravitational wave is polarized, and though the polarization is fixed in terms of $\gamma+\Gamma$,
it can be arbitrary. In addition there is a scalar wave, the trace. In the case when $e_3\neq 0$ the perturbation $\h{I}$ is not of Petrov
type $N$; the additional gravitational mode is a constraint mode.

\initiate
\section{The Poisson Energy and conservation laws}                    \label{vtjam}

We have associated a gravitational field to the noncommutative
structure with the map (\ref{map}). We would like to consider now this
structure as an effective field and estimate its 
energy-momentum (`Poisson energy').  We are confronted immediately with the choice of the position of the extra term
in the Einstein equations. We can place it on the right-hand side
and consider it on the same level as any matter source. We
can also keep it on the left-hand side and consider it as a noncommutative
modification of the curvature. First however we make some preliminary remarks about conservation laws.

From (\ref{vtj5}-\ref{vtj6}) for the Einstein tensor we obtain
\begin{equation}
G_{\beta\gamma} 
= -\tfrac 12\epr\Big(J_{0\zeta(\beta} 
e^\zeta e^\alpha \h{I}_{\gamma)\alpha} 
+ J_0{}^{\alpha\zeta}e_\zeta e_{(\beta}\h{I}_{\gamma)\alpha} 
- 2\eta_{\beta\gamma}J_{0\zeta\delta}
e^\zeta e_\alpha\h{I}^{\delta\alpha}\Big).              \label{Einst}
\end{equation} 
In general the Einstein tensor does not vanish.  A conservation equation of the associated energy-momentum tensor in  linear  approximation is easy to verify.  Applying the cocycle
condition and keeping in mind that, to linear order in $\epr$,
$e_\alpha e_\beta = e_\beta e_\alpha$, we obtain
\begin{eqnarray}
&& 
e^\beta G_{\beta\gamma} = -\tfrac 12\epr \Big(
J_{0}{}^{\alpha\zeta}e^\beta e_\zeta e_\gamma \h{I}_{\beta\alpha} 
+J_{0}{}^{\alpha\zeta}e^\beta e_\zeta e_\beta \h{I}_{\gamma\alpha} 
-2J_{0\zeta\delta}e_\gamma e^\zeta e_\alpha\h{I}^{\delta\alpha}\Big) 
\nonumber \\[8pt]&&\phantom{e^\beta G_{\beta\gamma}}
=- \tfrac 12\epr J_0{}^{\delta\zeta}
e_\zeta e^\alpha ( e_\alpha \h{I}_{\gamma\delta} 
-e_\gamma \h{I}_{\alpha\delta})
\nonumber \\[8pt]&& \phantom{e^\beta G_{\beta\gamma}}
= \tfrac 12\epr J_0{}_{\delta\zeta}e^\zeta e^\alpha e^\delta
\h{I}_{\alpha\gamma} = 0.
\end{eqnarray} 
As we shall see, the conservation law holds in an important special
case in quadratic order too.

\subsection{Canonical orientation}

To the extent that the noncommutative background is analogous to a
lattice, the perturbations can be considered as elastic vibrations or
phonons. This analogy however is tenuous at the approximation we are
considering since we have excluded any resonance phenomena. These
could appear if we allowed larger-amplitude waves with energy
sufficient to change the background. The case we shall now focus to
would then be analogous to a phonon propagating along one of the axes of a
regular cubic lattice. In the special case in which it is
also Petrov vector of the perturbation the dispersion relations become
clearer. 

Assume then that $\eta$ and $\xi$
are parallel and set 
\begin{equation}
\eta^\alpha = J_0^{\alpha\beta}\xi_\beta =\lambda \xi^\alpha.                       \label{cash}  
\end{equation}
It follows from~(\ref{ec})  that the vector $\xi$ is an
eigenvector of $J$ also to second order.
Equation~(\ref{R}) yields for the Riemann curvature tensor to linear order 
\begin{equation}
R_{\alpha\beta\gamma\delta} =0.
\end{equation} 
The dispersion relation
\begin{equation}
\xi^2 = 0
\end{equation}
follows from~(\ref{cyc2}).  

We stress that we have no action and that this dispersion relation was
not obtained from field equations. It is valid however only in the
case of wave propagation satisfying the relation~(\ref{cash}). There
is a certain amount of obscurity surrounding the role of the cocycle
condition, to what extent and how it can be used to replace the
field equations. Suppose for simplicity that the scalar field $\chi$ 
vanishes.  Using the dual object $\h{I}^*$ one can write the cocycle
condition as the condition
\begin{equation}
\h{I}^{*\alpha\beta} \xi_\beta = 0.
\end{equation}
On the other hand Equation~(\ref{ec}) in the particular orientation we
have chosen for the wave propagation is written as
\begin{equation}
\h{I}_{\alpha\beta}\xi^\beta = 0.        
\end{equation}
We can conclude then that for any complex $c$ the difference
\begin{equation}
H=\h{I}^{*} -c\h{I}  
\end{equation}
is orthogonal to the propagation vector. We are considering the WKB
approximation which implies that the essential direction is in fact
$\xi$. We can conclude then that `essentially' we have $H=0$. But this
is a modified self-duality condition, that is a condition which
equates a dynamical object with a topological one.

In quadratic order, using the dispersion relation, we find that the
expression~(\ref{Ric}) simplifies to
\begin{equation}
\vev{R_{\beta\gamma}} = - \tfrac 18\epr^2(i\omega)^2 
(\a{\chi}^2 + 2\lambda^2 \h{\a{I}}_{\alpha\eta}\h{\a{I}}^{\alpha\eta})
\xi_\beta\xi_\gamma
\end{equation}
for the Ricci tensor.  The corresponding expression for the Ricci
scalar vanishes and we obtain for the Einstein tensor the average value
\begin{equation}
  \vev{G_{\beta\gamma}} = -\rho \xi_\beta\xi_\gamma     \label{einsttens}
\end{equation}
with
\begin{equation}
\rho = -\tfrac 18(\epr\omega)^2 
(\a{\chi}^2 + 2\lambda^2 \h{\a{I}}_{\alpha\eta}\h{\a{I}}^{\alpha\eta}) .
\end{equation}
The energy-momentum is that of a null dust with a density $\rho$.

In the WKB approximation we can, just as in the classical case, derive
a conservation law for $\rho$ which has a natural interpretation as
graviton-number conservation. The conservation law however now can be
derived directly from the cocycle condition.  If we multiply the
cocycle condition~(\ref{dI}) by $\xi^\alpha$ we obtain
\begin{equation}
\xi^\alpha e_\alpha \h{I}_{\beta\gamma}
+ \xi^\alpha e_\beta \h{I}_{\gamma\alpha}
+ \xi^\alpha e_\gamma \h{I}_{\alpha\beta} = 0.       \label{2ndcyc}
\end{equation}
We also have
\begin{equation}
 e^\alpha(\xi_\alpha \h{I}_{\beta\gamma}
+ \xi_\beta \h{I}_{\gamma\alpha}
+ \xi_\gamma \h{I}_{\alpha\beta}) = 0.   
\end{equation}
Adding these two equations, using~(\ref{ec}), (\ref{cash}) and not 
forgetting that
$e_\alpha\xi_\beta = e_\beta\xi_\alpha$ in our approximation, 
we find
\begin{eqnarray}
&&
 (\xi^\alpha e_\alpha \h{I}_{\beta\gamma}
+ e^\alpha( \xi_\alpha \h{I}_{\beta\gamma})) \h{I}^{\beta\gamma}
+2 (\xi^\alpha e_\beta \h{I}_{\gamma\alpha}
+  e^\alpha( \xi_\beta \h{I}_{\gamma\alpha})) \h{I}^{\beta\gamma}
\nonumber\\[4pt]&&
=e_\alpha(\xi^\alpha\hI^{\beta\gamma }\hI_{\beta\gamma}) 
+ 2e^\alpha(\xi^\beta\hI^{\gamma\alpha}\hI_{\beta\gamma}) =0  .
\end{eqnarray}
The conservation law
\begin{equation}
e^\alpha (\rho \xi_\alpha) = 0
\end{equation}
follows and from it the conservation of the effective source,
\begin{equation}
e^\alpha(\rho \xi_\alpha\xi_\beta) =0.
\end{equation}

To interpret the additional term we have isolated
as the energy-momentum of an external field, 
\begin{equation}
G_{\beta\gamma} = -16\pi G_N  T_{\beta\gamma},
\end{equation}
the sign of $\rho$ should be non-negative.  However, as
\begin{equation}
\rho = -\tfrac 18(\epr\omega)^2 
(\a{\chi}^2 + 2\lambda^2 \h{\a{I}}_{\alpha\eta}\h{\a{I}}^{\alpha\eta})
= \tfrac 14(\epr\omega\lambda{\vec e})^2
- \tfrac 14(\epr\omega\lambda{\vec b})^2
 - \tfrac 18(\epr\omega\a{\chi})^2  ,
\end{equation}
the matter density does not have a fixed sign, unless of course one
place restrictions on the relative importance of the space-time and
space-space commutation relations. 
This exactly is one of the properties which can explain the 
acceleration of the universe~\cite{SahSta06} and it makes the `Poisson energy'  
a possible candidate for dark energy. We shall examine
this possibility in a subsequent article.

\subsection{Sparling's forms}

In the Cartan frame formalism the field equations are most elegantly
(and, once one is familiar with it, easily) written as the vanishing
of a 3-form. This follows from the identity
\begin{equation}
G_{\alpha\beta} \wm \theta^\beta
= - \tfrac 12 \Omega^{\beta\gamma} \wm \theta_{\alpha\beta\gamma} .
\end{equation}
We have here introduced
\begin{equation}
\wm \theta^\alpha = 
\frac1{3!} \epsilon^{\alpha\beta\gamma\delta}
\theta_\beta\theta_\gamma\theta_\delta
\end{equation}
as well as its inverse
\begin{equation}
\wm \theta^{\alpha\beta\gamma} 
= \epsilon^{\alpha\beta\gamma\delta}\theta_\delta.
\end{equation}
Since the signature is that of Minkowski we find that $\wm^2=-1$.

We write the energy-momentum of the gravitational field in
terms of a vector-valued 3-form  (`S' for Sparling)
\begin{equation}
\tau_{S\alpha} = \frac 12  \epsilon_{\alpha\beta\gamma\delta}
(\delta^\beta_\zeta\omega^{\gamma\eta} \omega_\eta{}^{\delta}
- \omega^{\gamma\delta}\omega^{\beta}{}_{\zeta}) \theta^\zeta
\end{equation}
which has the property~\cite{DubMad87} that it is exact if and only if
the Einstein field equations are satisfied. If this be so one sees
that the total energy-momentum is given as the integral over the
sphere at infinity of the Sparling 2-form
\begin{equation}
\sigma_\alpha = - \tfrac 12 \omega^*_{\alpha\beta}  \theta^\beta ,
\qquad \omega^*_{\alpha\beta} 
= \frac 12 \epsilon_{\alpha\beta\gamma\delta} \omega^{\gamma\delta}.
\end{equation}
Since in general there is a preferred frame, that canonically aligned
with respect to the eigenvectors of the conformal tensor, one can
claim that the 2-form itself and not only the integral thereof is
well-defined.

We can then also consider the extra terms we obtained in the previous
section to be due to a noncommutative extension of the curvature and
write Equation~(\ref{einsttens}) as a vacuum equation
\begin{equation}
\vev{G_{\beta\gamma}} +\rho \xi_\beta\xi_\gamma = 0.       \label{vac}
\end{equation}
The field equations~(\ref{einsttens}) can then be written as the
condition (`P' for Poisson)
\begin{equation}
G_{\alpha\beta}\wm \theta^\beta - \tau_{P\alpha} =  0.
\end{equation}
Classically one has the identity
\begin{equation}
G_{\alpha\beta}\wm \theta^\beta + \tau_{S\alpha} = d\sigma_\alpha .
\end{equation}
We can write then 
\begin{equation}
\tau_{P\alpha} + \tau_{S\alpha} = d\sigma_\alpha.
\end{equation}
The vacuum field equations~(\ref{vac}) are the integrability
conditions for the modified system
\begin{equation}
d\tau_{\alpha} = 0, \qquad 
\tau_{\alpha} = \tau_{P\alpha} + \tau_{S\alpha}.
\end{equation}
However, only when we have calculated the modification in a few more examples
can we hope to lift the change of the 3-form to a change of the curvature
form.

\initiate
\section{Conclusions} 

The formalism on which the article has been based is one with a
preferred frame. It is in a sense then gauge-fixed from the
beginning. We have shown that the degrees-of-freedom or basic modes of
the resulting theory of gravity can be put in  correspondence
with those of the noncommutative structure. As an application of the
formalism we have considered a high-frequency perturbation of the
metric.  In the classical theory it follows from the field equations
that the perturbation must satisfy a dispersion relation and a
conservation law. We show that these remain valid in the
noncommutative extension of the frame formalism and that they are 
consequences of a cocycle condition on
the corresponding perturbation of the Poisson structure. 

We have also shown that the perturbation of the Poisson structure
contributes to the energy-momentum as an additional effective source
of the gravitational field. Although the explicit form of this
contribution, the Poisson energy, was calculated only in a linearized,
high-frequency approximation it is certainly
significant in a more general context. We stress that because of the
identification of the gravitational field with the Poisson structure
the perturbation of the latter is in fact a reinterpretation of a
perturbation of the former and not an extra field. The difference with
classical gravity lies in the choice of field equations. In the WKB
approximation  this amounts only to a modification of
the conserved quantity.

\initiate
\section*{Acknowledgment}
This work is supported by the EPEAEK programme "Pythagoras II" and
co-funded by the European Union(75\%) and the Hellenic state (25\%)
and by the Grant 141036 of MNTR, Serbia.


\begin{thebibliography}{10}

\bibitem{Pau56}
W.~Pauli, ``{R}elativit{\"a}tstheorie und {W}issenschaft'', {\em Helv.\ Phys.\
  Acta} {\bf 29 {Suppl IV}} 282--286.
\newblock Bern, July 1955.

\bibitem{Mad00c}
J.~Madore, {\em An Introduction to Noncommutative Differential Geometry and its
  Physical Applications}.
\newblock No.~257 in London Mathematical Society Lecture Note Series. Cambridge
  University Press, second~ed., 2000.
\newblock 2nd revised printing.

\bibitem{BurMad05b}
M.~Buri{\'c} and J.~Madore, ``A dynamical 2-dimensional fuzzy space'', {\em
  Phys.\ Lett.} {\bf B622} (2005) 183--191,
\href{http://xxx.lanl.gov/abs/hep-th/0507064}{{\tt hep-th/0507064}}.

\bibitem{Isa68a}
R.~A. Isaacson, ``Gravitational radiation in the limit of high frequency. {I}.
  {T}he linear approximation and geometrical optics'', {\em Phys.\ Rev.} {\bf
  160} (1968) 1263---1271.

\bibitem{Isa68b}
R.~A. Isaacson, ``Gravitational radiation in the limit of high frequency. {II}.
  {N}onlinear terms and the effective stress tensor'', {\em Phys.\ Rev.} {\bf
  160} (1968) 1272---1280.

\bibitem{Mad73}
J.~Madore, ``The absorption of gravitational radiation by a dissipative
  fluid'', {\em Commun.\ Math.\ Phys.} {\bf 30} (1973) 335.

\bibitem{BurGraMadZou06a}
M.~Buri{\'c}, T.~Grammatikopoulos, J.~Madore, and G.~Zoupanos, ``Gravity and
  the structure of noncommutative algebras'', {\em J.~High Energy Phys.} {\bf
  04} (2006) 054--070,
\href{http://xxx.lanl.gov/abs/hep-th/0603044}{{\tt hep-th/0603044}}.

\bibitem{Mad97c}
J.~Madore, ``On {P}oisson structure and curvature'', in {\em Coherent states,
  differential and quantum geometry}, S.~T. Ali, A.~Odzijewicz,
  M.~Schlichenmaier, and A.~Strasburger, eds., vol.~43 of {\em Rep.\ on\ Math.\
  Phys.}, pp.~231--238.
\newblock 1999.
\newblock \href{http://xxx.lanl.gov/abs/gr-qc/9705083}{{\tt gr-qc/9705083}}.
\newblock
{B}ialowieza, July 1997.

\bibitem{Sch99}
V.~Schomerus, ``{$D$}-branes and deformation quantization'', {\em J.~High
  Energy Phys.} {\bf 06} (1999) 030,
  \href{http://xxx.lanl.gov/abs/hep-th/9903205}{{\tt hep-th/9903205}}.

\bibitem{AleRecSch00}
A.~Y. Alekseev, A.~Recknagel, and V.~Schomerus, ``Brane dynamics in background
  fluxes and non-commutative geometry'', {\em J.~High Energy Phys.} {\bf 05}
  (2000) 010,
\href{http://xxx.lanl.gov/abs/hep-th/0003187}{{\tt hep-th/0003187}}.

\bibitem{Ste07}
H.~Steinacker, ``Emergent gravity from noncommutative gauge theory'',
\href{http://xxx.lanl.gov/abs/arXiv:0708.2426 [hep-th]}{{\tt arXiv:0708.2426
  [hep-th]}}.

\bibitem{DubKerMad89b}
M.~Dubois-Violette, R.~Kerner, and J.~Madore, ``Classical bosons in a
  noncommutative geometry'', {\em Class.\ and Quant.\ Grav.} {\bf 6} (1989),
  no.~11,
1709--1724.

\bibitem{MadSchSchWes00a}
J.~Madore, S.~Schraml, P.~Schupp, and J.~Wess, ``Gauge theory on noncommutative
  spaces'', {\em Euro.\ Phys.\ Jour.~C} {\bf 16} (2000) 161--167,
  \href{http://dx.doi.org/10.1007/s100520000394}{{\tt
  DOI:10.1007/s100520000394}},
\href{http://xxx.lanl.gov/abs/hep-th/0001203}{{\tt hep-th/0001203}}.

\bibitem{MadSchSchWes00b}
J.~Madore, S.~Schraml, P.~Schupp, and J.~Wess, ``External fields as intrinsic
  geometry'', {\em Euro.\ Phys.\ Jour.~C} {\bf 18} (2001) 785--794,
\href{http://xxx.lanl.gov/abs/hep-th/0009230}{{\tt hep-th/0009230}}.

\bibitem{DubMad87}
M.~Dubois-Violette and J.~Madore, ``Conservation laws and integrability
  conditions for gravitational and {Y}ang-{M}ills field equations'', {\em
  Commun.\ Math.\ Phys.} {\bf 108} (1987)
213--223.

\bibitem{SahSta06}
V.~Sahni and A.~Starobinsky, ``Reconstructing dark energy,''
 {\em Int.\ J.\ Mod.\ Phys.\  D} {\bf 15} (2006) 2105-2132
 \href{http://xxx.lanl.gov/abs/astro-ph/0610026},{{\tt astro-ph/0610026}}.



\end{thebibliography}
\providecommand{\href}[2]{#2}\begingroup\raggedright\endgroup
\end{document}